%% file: main.tex
\documentclass[runningheads]{llncs}
\usepackage[T1]{fontenc}
\usepackage{graphicx}
\usepackage{hyperref}
\usepackage{color}

\usepackage[utf8]{inputenc}
\usepackage[normalem]{ulem}
\usepackage{comment}
 
\begin{document}
\title{Perils of current DAO governance} 
%
%
\author{Aida Manzano Kharman\inst{1, 3}\orcidID{0000-0002-5342-3037} \and
Ben Smyth\inst{2,3}\orcidID{0000-0001-5889-7541} 
}
\authorrunning{Manzano Kharman and Smyth.}
%
\institute{Imperial College London, UK \and
University of Birmingham, UK \and
VoteTech Ltd, UK \\
\email{amm3117@ic.ac.uk}\\ \email{io@bensmyth.com}\\
}
\maketitle              
\begin{abstract}
DAO Governance is currently broken. We survey the state of the art and find worrying conclusions. Vote buying, vote selling and coercion are easy. The wealthy rule, decentralisation is a myth. Hostile take-overs are incentivised. Ballot secrecy is non-existent or short lived, despite being a human right. Verifiablity is achieved at the expense of privacy. These privacy concerns are highlighted with case study analyses of Vocdoni's governance protocol. This work presents two contributions: firstly a review of current DAO governance protocols, and secondly, an illustration of their vulnerabilities, showcasing the privacy and security threats these entail.

\end{abstract}

\input{Intro2.tex}

\input{DAO_Governance_Fact_or_Fiction.tex}

\input{LiteratureReview.tex}
\input{KeyPlayers.tex}
\input{Conclusion.tex}

\subsubsection{Acknowledgements} Aida Manzano Kharman acknowledges and thanks IOTA Foundation for the funding of her PhD studies.

%
%
%
%
\bibliographystyle{splncs04}
\bibliography{mybibliography}

\end{document}

%% file: Intro2.tex
\section{Introduction}

Welcome to Web3: The era of quick riches~\cite{murray2021sell,BELK2022198}. Everyone wants a slice, especially since they realised they are the pie~\cite{potts2019web3,gilder2018life,al2012data}. Gone are the days where the users provide value and the services reap the reward~\cite{schneble2020google,berghel2018malice,nabben2021DAO,Datacy}. Users want a voice and a share of the reward~\cite{serada2022economy}. 
As a result, an online revolution is unfolding. 

Web3's paradigm shift is not new. 
For centuries collectives have organised to redistribute centralised power and create a democracy~\cite{wade2017russian,carlyle2019french}. 
They sought control, a say in their future, lives and income.
A DAO\footnote{Decentralised Autonomous Organisations} enables shared decision making amongst netizens~\cite{merkle2016daos}. Users actively control services in which they participate~\cite{sims2019blockchain,bellavitis2022rise}. But do they? We uncover the truth: Wealthy minorities amass voting power, vote buying is legal, vote selling is incentivised, coercion is easy. We dig into the hows and the whys and illustrate these weaknesses with a case study on Vocdoni's governance platform.

%% file: DAO_Governance_Fact_or_Fiction.tex
\section{DAO Governance: Fact or Fiction?}
\newcommand{\HRule}{\rule{\linewidth}{0.2mm}}
It's 2016: 
DAOs are in their infancy, 
The DAO\footnote{Confusingly, \emph{The DAO} is the name of a DAO.} has garnered attention having raised \$150 million of Ethereum tokens.
Three months after launch, The DAO is hacked, a smart contract bug exploited,~\cite{theDAOhack}
\$50 million siphoned off their funds~\cite{DAOHACK}. 
The aftermath raising questions over blockchain philosophy and the technology's future.

Were funds obtained legally?
`Code is law' is regulation enforced by technology~\cite{hassan2017expansion}. 
It underpins the functioning of DAOs and blockchain.
If software is exploitable, no law is broken. 
Victims lost their funds unfairly.
Ethereum founder Vitalik Buterin proposed a soft-fork (a software update in the blockchain proposal)
to right the `wrong'.
The solution was promptly abandoned; it too contained a bug, making it vulnerable to further attacks.

The tokens amassed by the attacker gave them enough legislative power to enact decisions in The DAO.
The alleged attacker responded by threatening to bribe miners to not comply with the soft-fork. 
They argued no smart contract rules were broken when obtaining the funds. 

The DAO's value exceeded the cost of acquiring enough votes to take control, incentivising `the heist.' 
There is no need to break the laws established by the DAO to succeed.

Fast-forward to 2018: History repeats, another DAO is victim to poor governance. 
This time no bug was exploited, the attacker simply acquired enough tokens,
bought the vote, approved their own proposal. 
The coup drained nearly \$500,000 tokens from the Build Finance DAO.\footnote{%
\href{https://www.vice.com/amp/en/article/xgd5wq/democratic-dao-suffers-coup-new-leader-steals-everything}{DAO Coup, Vice}}
The attacker covered their tracks using Tornado Cash, anonymising stolen funds. 
Token-based voting legalises coups---anyone can legitimately buy their way to power.
Incentive makes takeovers inevitable if the cost is cheaper than the reward.

Democracies embrace one-person one vote. 
Acquiring multiple votes undermines fairness, equality. 
Token-based voting is incompatible with equality and fairness. 
Tokens are not a proxy for identity, their ownership is easily transferred. 
Wealth amasses tokens, buys legislative power, corrupts decision making~\cite{hostile}. 
A voting system that allows voters to buy more votes converges to plutocracy, the unwanted symptoms of centralisation, low representation of the electorate~\cite{barbereau2022decentralised} and game theoretic incentives to attack the DAO~\cite{plutocracy}.\\  
\HRule \\
{\textbf{Sidebar1: Public Votes and Vote Selling}}\\
\HRule \\
Game theory allows for a better understanding of vote selling.
Wealthy agents buy voting power. 
When it comes to voting, small to mid-sized token holder's votes are not as powerful. 
In an election, there is no incentive for them to vote against the wealthy agents, because to cast a vote on-chain, voters must also pay a transaction fee.
Voter's are economically incentivised to abstain~\cite{mark2016call}!
Worse---voters are economically incentivised to sell their vote for financial reward. 
The latter is always a winning strategy.  

A terrifyingly simple proposition: Rationale vote buyers can confirm their purchases.  
Votes are typically revealed during or after an election, compliance can be verified.
Secondly, the ownership transfer of a vote is remarkably easy. 
The voting ability and power is linked strictly to tokens, not to an identity. Crypto-currencies enable fast and simple transfer of said tokens.
Vote-buying cartels can emerge: From simple smart contracts to pay out voters automatically upon proving compliance, to cartels buying trusted hardware executing vote buying code\cite{austgen2023dao}. 
Particularly, the latter is an attack vector that is essentially undetectable~\cite{daian2018chain}.

The insights gathered in~\cite{mosley2022dash} confirm the incentive to abstain, the dangers of public voting and the centralisation of power. 
DAO governance was studied with a focus on Dash DAO as a case study.
Researchers accessed the voting history of Dash DAO's masternodes, given that these are public. 
Worryingly, IP identifiers, software version and wallet addresses were public too.
Voting patterns of 4987 masternodes who participated in voting across 577 proposals were analysed. Researchers found that: `Some masternodes are not only abstaining from
voting, but have disengaged from the voting process completely.'~\cite{mosley2022dash}.
They also found a number of voters with almost identical IP Ports, strongly indicating that they are mounting sybil attacks to gain voting power.
Further to this, they analysed the voting patterns of the DAO participants. 
Results show that there are small-sized, dense clusters of masternodes with identical voting patterns. 
Although smaller in number compared to the rest of voters, if these minority clusters were to collude, `they would have more voting power than the entire decentralised majority'~\cite{mosley2022dash}. 
\HRule\\
Vote buying, public votes and paying to vote are the harsh reality of DAO governance. 
The consequences: low turnout, centralisation, preclusion of free will, coups and coercion. 
A preliminary study found less than 1\% of token holders control 90\% of the vote~\cite{chainalysis}.\footnote{
Chainalysis only studied ten DAOs, further study would establish general trends.}
Are DAOs decentralised when controlled by a wealthy minority? 
Clearly not---the wealthy do not represent the masses.

%% file: LiteratureReview.tex
\newcommand{\BS}{\fontfamily{lmss}\selectfont Ballot Secrecy}
\section{My Vote: My Business}

Historically, \textit{``Americans [voted] with their voices -- \emph{viva voce} -- or with their hands or with their feet. Yea or nay. Raise your hand. All in favor of Jones, stand on this side of the town common; if you support Smith, line up over there"}~\cite{lepore2008rock}. 
Everyone present could verify that only voters voted and that the count was correct. 
But free will must be ensured, as dictated by the United Nations \cite{assembly1948universal}, the Organisation for Security \& Cooperation in Europe~\cite{osce}, and the Organization of American States~\cite{OAS}.
Yet public votes forgo freedom, ``The unfortunate voter is in the power of some opulent man; the opulent man informs him how he must vote. 
Conscience, virtue, moral obligation, religion, all cry to him, that he ought to consult his own judgement, and faithfully follow its dictates. 
The consequences of pleasing, or offending the opulent man, stare him in the face...the moral obligation is disregarded, a faithless, ..., pernicious vote is given''~\cite{mill}.
The need for voting privately became evident. 
In-person voting ensures this by providing identical ballots that are completed in a private booth, a concept first introduced successfully in Australian voting in 1856~\cite{newman2003tasmania}. 
\HRule \\
{\textbf{Sidebar2: Ballot Secrecy in e-voting}}\\
\HRule \\
In e-voting, the concept of secret ballots emerged parallel to the development of such voting schemes, originating with David Chaum's first proposal of an end-to-end verifiable voting scheme in 1981. 
In it, voter's ballots were private, and all participants could check that the tallying operation was correctly performed \cite{chaum1981untraceable}.
\HRule\\

Forgoing ballot secrecy is to regress centuries of progress, violate human rights and returning coercion and inequality as norms. 
With that in mind, we warn: DAOs are in dire straits...

\section{DAO Voting: Survival of the Richest}

DAO members vote remotely, online. 
One of the methods is on-chain voting, where the public nature of distributed ledgers is leveraged, using them as a shared and verifiable database. 
Proposals are encoded into smart contracts and submitted to the ledger as a transaction. 
A vote in favour or against new proposals is cast as a transaction on the ledger. 
Winning proposals are automatically executed. Votes, proposals and election outcomes are all publicly verifiable~\cite{muth2021empirical,lee2016electronic}. 
On-chain voting makes elections outcomes binding, without relying on a trusted intermediary or a board to implement results.
Guarantees of immutability are provided by the ledger: Once the results are announced, these cannot be tampered with. Mounting an attack to re-write the blocks requires practically infeasible computational power. 
On-chain governance uses distributed ledgers as a public (or permissioned, depending on the protocol) bulletin board. 
Despite its desirable properties, it has been subject to criticism~\cite{8405627,tacs2020systematic}. Its detractors argue that blockchain voting not only fails to mitigate security risks present in e-voting, but also introduces additional risks~\cite{park2021going}. We agree. 

Worryingly, the vast majority of on-chain, smart contract votes do not satisfy ballot secrecy. 
At worst, votes are revealed as cast, and at best, these are publicly decrypted after the voting period ends. 
Values of a token can be artificially inflated or devalued,
`pump and dumps' become simple.
Whales (entities or individuals with large amounts of tokens) can manipulate the value of a token with their behaviour.
They can express intention with public votes, swaying token values to their favour. 
Just before the election closes, they change their intention, make a profit and cash out.

Information on how a wallet address voted, when, and how many tokens they staked to that vote is available for anyone in the ledger to see. 
Wallet addresses are pseudonymous, not anonymous~\cite{androulaki2013evaluating}, and it is possible to link wallet addresses to individuals from information such as their transaction history~\cite{biryukov2015bitcoin}.
Tornado Cash hides this, but has also been maliciously used to launder millions---the U.S. Department of the Treasury’s Office of Foreign Assets Control (OFAC) recently sanctioned the crypto-currency mixer\footnote{\href{https://www.cnbctv18.com/cryptocurrency/tornado-cash-the-coin-mixer-\\sanctioned-by-us-treasury-for-allegedly-laundering-7b-worth-virtual-currency-14430572.htm}{Tornado Cash Sanctioned, CNBC}} and the developers were arrested.\footnote{\href{https://thehackernews.com/2022/08/tornado-cash-developer-arrested-after.html}{Torndado Cash Developers Arrested, The Hacker News}}
On-chain transaction fees means voters pay to vote. Fees soar unpredictably, unfairly discriminating between voters. 
They can be victims of miners refusing to cast their votes, and only the wealthiest will survive the financial hurdles.
Paying to vote or increasing the weight of their vote proportional to their wealth discriminates against those who cannot do so from the decision making process. 
What if a coup happens? Forking the chain brings little solace: election records can be reverted, actual events cannot, history cannot be changed; assets may have already been cashed out.

\subsection{Off-Chain Voting and Hybrid Alternatives}

Alternatives exist that don't use blockchain to cast votes. 
The most popular example is Snapshot, which many DAOs use solely or in combination with on-chain voting to enable governance. 
Snapshot is decentralised, using IPFS as its main storage layer~\cite{IPFSDocs}.
It offers the advantage of no fees to cast a vote whilst still being decentralised thanks to their storage system. 
The election outcome however, is not automatically binding. It has to be bought on-chain.
Because of this, Snapshot is often used for polling. 
AragonDAO, Uniswap and MakerDAO are examples of DAOs using a hybrid governance solution~\cite{aragon,adams2021uniswap,makerdao}. 

 

%% file: KeyPlayers.tex
\section{A New Hope?}

Despite the dire situation of DAO governance, we observe that a shy but steady shift is occurring in the space.
A number of projects are emerging to address some of the aforementioned issues, although they are still in their infancy. 

Snapshot is pairing with Orange Protocol to develop a reputation based voting mechanism~\cite{OrangeDocs}.
Responding to inequality, communities such as Algorand~\cite{algorand} and Dream DAO~\cite{dreamDAOgovernance} are transitioning towards a merit based voting system to actively encourage participation and development of the network, and distribute voting power amongst the developers, not the wealthy.
Moving away from vote purchasing governance models is necessary to avoid plutocracies and centralisation and `legal' fund siphoning.

\begin{figure} 
\includegraphics[width=\textwidth]{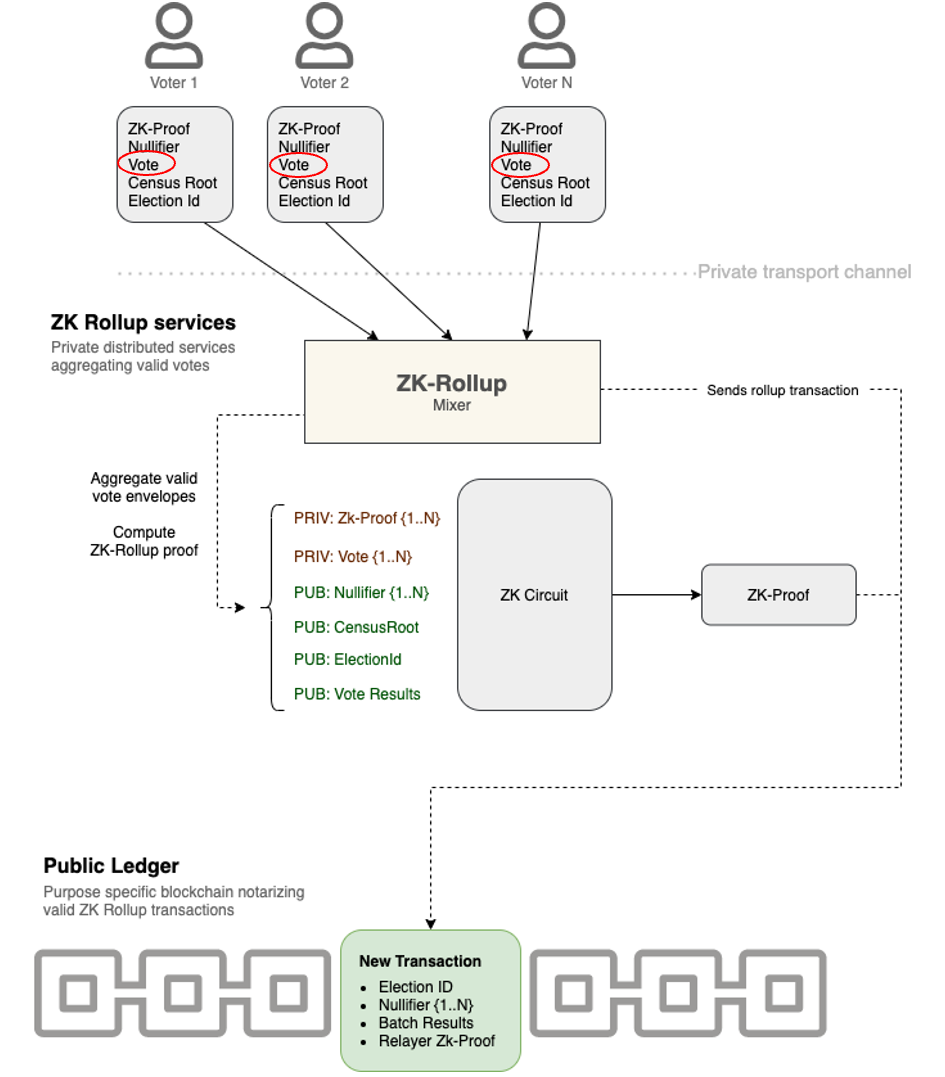}
\caption{Vocdoni's Zero Knowledge Rollup Proposal. Source: \href{https://docs.vocdoni.io/architecture/protocol/rollup.html}{Vocdoni}} 
\label{VocdoniZK}
\end{figure}

To address ballot secrecy, VoteCoin presents an on-chain voting solution offering encrypted ballots during the election process~\cite{VoteCoin}. 
Snapshot are also developing a similar feature, offering `shielded voting' whereby votes are private only until the end of the election.\footnote{\href{https://decrypt.co/105201/snapshot-adds-shielded-voting-daos-help-solve-voter-apathy}{Snapshot shielded voting}}
Privacy in this case, is short lived.
A number of issues remain: verifiability is achieved at the expense of privacy by naively decrypting votes publicly. 
An option exists to allow an auditor to decrypt votes, but this introduces a trust assumption of honesty of the auditor. 
VoteCoin also requires voters to pay to cast their ballot.
A promising on-chain voting protocol is MACI \cite{MACIG}. In it, voters encrypt their votes and a trusted coordinator is tasked with decrypting the ballots and returning an election outcome. This scheme introduces a strong trust assumption: the coordinator must indeed be trustworthy, as they have the power to decrypt individual ballots and therefore know how each voter voted. This protocol does not satisfy formal notions of ballot secrecy as defined in \cite{smyth2021ballot}.
Another relevant case study is Aragon DAO's new governance solution: Vocdoni. 
They provide an on-chain voting solution that uses two blockchains: the Ethereum blockchain for the election process creation or status update, and the Vochain blockchain (Vochain), where votes are cast \cite{vocdoni2021}. 
Vochain uses the Proof of Authority Tendermint blockchain, so only trusted nodes can relay transactions. 
Due to the use of two blockchains, there is a need for an oracle to relay information from the Ethereum blockchain to Vochain, to signal new voting processes. 
At time of writing, the oracle nodes are run as trusted nodes, however, Vocdoni proposes a roadmap to substitute them with Zero-Knowledge Rollups\footnote{A Zero-Knowledge Rollup is a proof system used to compress a number of transactions into a batch, with cryptographic assurance that these are correct. A more detailed overview is presented in~\cite{augot2022zero}.} to allegedly make them trustless. 
According to Vocdoni's documentation: “One solution to this problem is to make use of Zero-Knowledge Rollups as a vote aggregation and mixing mechanism between voters and the blockchain. 
This would make it impossible for any third party to verify that a voter chose a specific option”~\cite{vocdoni2021}. 
This claim is incorrect. As shown in Figure \ref{VocdoniZK}, the node computing the Zero-Knowledge Rollup receives the vote unencrypted, so they must be a trusted node. If this is not the case, the node computing the Zero-Knowledge Rollup can very easily reveal how a user voted. While the voter ID remains private, the prover computing the Zero-Knowledge Rollup will still know how a voter voted, given that it is them who send the vote to the prover in the first place. 
Even if the identity that a voter provides is a wallet address, these are pseudonymous. Indeed, the only obfuscated information is the ID of the voter within the census. 
Instead, the voter sends a zero-knowledge proof\footnote{A zero-knowledge proof is a way to prove that someone knows a piece of information without having to reveal it \cite{goldreich1994definitions}} of inclusion demonstrating that their ID belongs to the set of accepted voters. 

To understand the implications of this, we illustrate a parallel example for the reader: On the day of voting, anyone wishing to vote must cast their vote publicly, but what is hidden is their ID card, replaced instead by a proof that you hold a valid ID card and thus should be allowed to vote. 
Nonetheless, the voter must hand their public votes to the administrators, who can easily see how they voted, and could identify them, because indeed you were the one to hand them your vote. 
This implies that first, a great deal of trust must be placed on the administrators to not reveal your vote to malicious agents, and second, that no one else except the administrators will be able to observe your ballot as you cast it.  
Vocdoni addresses the second assumption by mentioning that a private transport channel would be used to send the votes to the prover. 
This assumption introduces a weaker notion of security, and the fact that the votes remain public in this channel means that this system cannot provide notions of ballot secrecy wherein the adversary is assumed to have the power of intercepting ballots during their collection. 
We would like to highlight that ballot secrecy does not equate to public votes with anonymous identities. 
Furthermore, the identities are not anonymous in Vocdoni, they are at best pseudonymous to the Zero-Knowledge Rollup prover, if the private transport channel is not compromised, and even making this assumption, voters would not be equal: later voters have more information with public votes. This is because Vocdoni does not support encrypted ballots with anonymous voting. 

We outline another vulnerability related to the ‘self-sovereign’ identity management of Vocdoni. In their protocol every user creates their own key pair \cite{vocdoni2021}. What is preventing users from selling their private key? In the anonymous voting, what is being hidden is the identity of the voter, and not their vote, so giving the voter the ability to generate their own identity would be parallel to allowing voters to create their own ID cards at an election. 
Instead of selling their vote, voters can sell their proof of census inclusion, that is directly generated from their identity. 
In fact, anyone can verify if this proof is invalid, so malicious agents attempting to coerce voters could easily check if they are being deceived. 
Similar to the Dark DAO vote buying cartels outlined by~\cite{daian2018chain}, identity buying cartels could emerge operating in the same manner. 
Indeed, black markets selling various types of identities already exist~\cite{USDOJ,stolenIDs}.

\begin{figure}
\includegraphics[width=\textwidth]{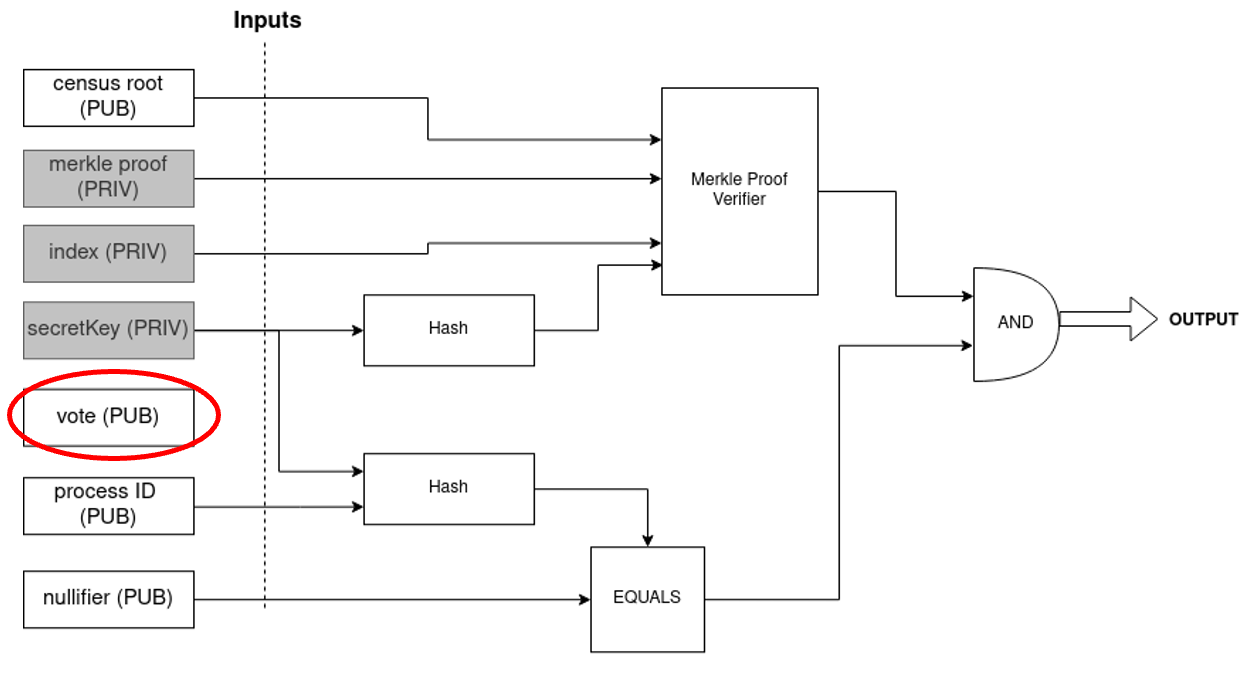}
\caption{Vocdoni's Anonymous Voting Schema. Source: \href{https://docs.vocdoni.io/architecture/protocol/anonymous-voting/zk-census-proof.html\#protocol-design}{Vocdoni}} \label{VocdoniAnon}
\end{figure}

Vocdoni does provide the option of having encrypted votes, but the voter identity remains known. They do not currently support both anonymous and encrypted voting at the same time.
Similar to VoteCoin and Snapshot's proposals, verifiability is once again achieved at the cost of privacy by publicly decrypting the results. 
With Vocdoni's anonymous voting, the ballots are public, as shown in Figure \ref{VocdoniAnon}. 
We again reiterate that anonymous voting with public votes does not achieve {\BS}. 

We summarise the state-of-the-art solutions in Table \ref{tab1}. The most used solution is on-chain smart contracts. 
It is a convenient option thanks to the existing integration platforms such as Tally and Boardroom providing a user-friendly platform to cast votes, tally them and summarise election outcomes.
No options provide long term {\BS}. 
Voter's identities are rarely kept private and verifiability comes at the expense of privacy in most. 

\begin{table}
\caption{Current DAO governance solutions.}\label{tab1}
\begin{tabular}{|p{2.9cm}|p{2cm}|p{2cm}|p{2.8cm}|p{2cm}|}
\hline
Voting Solution & {\BS} satisfied? &  Private Voter ID? & Naive Verifiability & Fee to cast Vote?\\
\hline
Snapshot &  No & No & Yes & No\\
Vocdoni &  No & Sometimes & Yes & Yes\\
VoteCoin & Temporarily & No & Yes & Yes\\
On-chain votes & No & No & Yes & Yes\\
MACI & No & Yes & No & Yes\\
\hline
\end{tabular}
\end{table}


%% file: Conclusion.tex
\section{Conclusion}

Since their birth in 2016, the emergence of DAOs has only but increased. This increase does not show any signs of slowing down. According to the data provided by DeepDAO~\cite{DeepDAO}, where in 2018 there were 10 DAOs, by 2020 there were approximately 200~\cite{forbes}. 
The influence and assets that DAOs hold has also increased. In 2021, the total Assets Under Management held by DAOs was \$520.7 million. Currently it has exploded to \$29.5 billion as of January 2024~\cite{DeepDAO}. Of particular importance is the value that these DAOs hold in their treasuries, which according to~\cite{slavin2022decentralized} had allegedly skyrocketed in 2021, from \$400 million to \$16 billion. Likewise, the number of DAO participants increased by 130 times from 13,000 to 1.6 million. 

We are witnessing a paradigm shift. With this explosion, a number of DAO projects have catastrophically crashed~\cite{morrison2020dao}. Hacks, scams, pump and dumps are rife~\cite{chain-crime-report}. The amount of value that has irreparably been lost as a consequence is humbling. We call for DAO practitioners to understand the risk that poor governance models entail. These are responsible for a number of DAO crashes. Flawed models put a target on the treasuries of vulnerable DAOs. Rationale actors will follow incentives: if the incentive to heist exists, DAOs cannot rely on the moral virtuousness of actors. Especially if many of these projects purport the narrative that ‘code is law’. 

The instances wherein an attacker acquires sufficient voting power to siphon treasury funds are not anecdotal\footnote{\href{https://decrypt.co/92970/build-finance-dao-falls-to-governance-takeover}{Build DAO's hostile governance takeover attack, Feb 2022}}, \footnote{\href{https://www.theverge.com/2022/4/18/23030754/beanstalk-cryptocurrency-hack-182-million-dao-voting}{Beanstalk cryptocurrency project robbed after hacker votes to send themselves \$182 million}}, \footnote{\href{https://www.bloomberg.com/news/articles/2023-05-21/sanctioned-crypto-mixer-tornado-cash-hijacked-by-hackers}{Sanctioned Tornado Cash DAO governance heisted by hacker}}. Mounting these heists are enabled by two core components, aside from poor governance models: flash loans and cryptocurrency mixers. Flash loans are defined as: \textit{`loans written in smart contracts that enable participants to quickly borrow funds without the need for collateral. These loans must be repaid in full within the same transaction, or else the entire transaction, including the loan itself, will be reversed.'}~\cite{eulerhack}. In the case of the Beanstalk DAO hack, the attacker emptied the DAO treasury using a flash loan, completing their attack in 13 seconds. They made an \$80 million profit. Subsequently, they anonymised the tainted transactions using Tornado Cash, an infamous cryptocurrency mixer. Funds were irreparably lost.  Although as mentioned earlier, Tornado Cash has been sanctioned by the OFAC, this does not bode the end for all crypto-currency mixers. Indeed, one of the architects of Tornado Cash is already working on an alternative: Privacy Pools~\cite{privacypool}. 
Flash loans are enabled by many platforms, examples include Aave~\cite{aave-docs}, and will continue to exist. The same can be said about crypto-currency mixers. Their underlying technology is open source. To prevent heist attacks, DAOs must ensure that their governance system is not exploitable.

Aside from the incentive to ward off hostile take-overs, good governance must be forefront in DAO agendas for the following reasons:
\begin{enumerate}
    \item It ensures the `Decentralised' adjective in the DAOs name actually holds true.
    \item It lays the cornerstone to have a flexible, democratic and updateable organisation.
    \item It provides provable security properties: with truly private votes, vote buying is prevented. Decisions are fair and free. 
\end{enumerate}

DAOs failing to provide these properties run the inevitable risk that sooner or later, an individual will follow incentives and empty their funds. Is that the fate DAOs are willing to accept?